# Rigorous Derivations of the Planar Impact Dynamics Equations in the Center-of-Mass Frame

Bob J. Scurlock, Ph.D., ACTAR and James R. Ipser, Ph.D.
*Department of Physics, University of Florida, Gainesville, Florida*

**Introduction**

In this article, we present a set of rigorous derivations of the equations used in planar impact dynamics calculations, in the context of vehicular accident reconstruction. We will provide an example application involving a collision between two passenger vehicles. We will also show how monte carlo analyses can be performed within the general framework provided by the equations. We also perform comparisons between energy loss functions for the approximate and exact solutions to the equations.

The reader is strongly encouraged to consult Brach's books on mechanical impact dynamics [1] and accident reconstruction [2] for a thorough treatment of the topic of planar impact dynamics. In addition, the reader may find it useful to consult [3] and [4] for additional aid.

**Relative Velocity Near Point-of-Contact**

In order to derive the standard equations needed to describe planar impact dynamics, we first start by defining the total velocity at any given point $P$ on or within vehicle $k$. We obtain this by first noting the position of $P$ can be expresses as the vector sum:

$$\bar{R}_k^P = \bar{R}_k^{CG} + \bar{r}_k^P$$

where $\bar{R}_k^P$ is the Earth-frame location of $P$, $\bar{R}_k^{CG}$ is the vehicle $k$ center-of-gravity (CG), and $\bar{r}_k^P$ is the location of $P$ measured with respect to the vehicle CG. Let us now assume that while point $P$ may be free to rotate with respect to the CG, the magnitude of $\bar{r}_k^P$ remains constant with time. In this case, taking the time derivative of both sides above, we obtain the expression:

$$\bar{v}_k^P = \bar{v}_k^{CG} + \bar{\omega}_k \times \bar{r}_k \quad (1)$$

where $\bar{v}_k^{CG}$ is the velocity at the center-of-gravity of vehicle $k$. $\bar{\omega}_k$ is the angular velocity of vehicle $k$ measured about its center-of-gravity, and $\bar{r}_k$ is the position vector extending from vehicle $k$'s CG to point $P$.

Next we need to define the relative velocity at or near the effective point-of-contact $PC$ between two vehicles undergoing collision. This is given by:

$$\bar{v}_{Rel}^{PC} = \bar{v}_1^{PC} - \bar{v}_2^{PC} \quad (2)$$

where the subscript $Rel$ will indicate the relative velocity between the two bodies – that is, the velocity of one vehicle with respect to the other. The superscript $PC$ will indicate at or near the effective point-of-contact.

Now that we have an expression for the relative velocity vector near the point-of-contact between the two vehicles, it will also be helpful to define the change in this vector due to impact forces. This is given by:

$$\Delta \bar{v}_{Rel}^{PC} = \bar{v}_{Rel,f}^{PC} - \bar{v}_{Rel,i}^{PC} \quad (3)$$

where the subscript $i$ will indicate the quantity immediately prior to impact and the subscript $f$ will indicate immediately after impact.

**Decomposition of Impulse**

From Newton's 3rd law, we know the collision forces exchanged by the two vehicles undergoing impact are equal in magnitude and opposite in direction. For simplicity, we assume that all other forces can be neglected during impact and that the collision force is constrained to be within the $x$-$y$ plane, where the $z$-axis points vertically upward. Let us assume the impact occurs on a flat roadway, where the road surface is parallel with the $x$-$y$ plane.

Let us also assume the collision force on vehicle 1 can be decomposed into "normal" and "tangent" components, where the normal force component can be associated a restoring force such as a linear spring force and the tangent component can be associated with frictional effects. Let $\hat{n}$ be the normal axis defining the direction of the restoring force and $\hat{t}$ be the tangent axis defining the direction of the frictional effects. We define a right-handed system such that $\hat{x} \cdot \hat{y} = 0$, $\hat{x} \times \hat{y} = \hat{z}$, $\hat{t} \cdot \hat{n} = 0$, and $\hat{t} \times \hat{n} = \hat{z}$ (see Figure 1).

Now we can write the collision force vector on vehicle 1 as:

$$\bar{F} = F_n \hat{n} + F_t \hat{t} \quad (4)$$

We use the subscript $n$ to identify the normal component of a vector, and $t$ to identify the tangent component. For an impact of duration given by $\Delta t$, the impulse on vehicle 1 is given by:

$$\bar{J} = \int_0^{\Delta t} dt \cdot \bar{F} = m_1 \Delta \bar{v}_1^{CG}$$
$$= \int_0^{\Delta t} dt \cdot F_n \hat{n} + \int_0^{\Delta t} dt \cdot F_t \hat{t} \quad (5)$$

where we identify the normal and tangent impulse components:

$$J_n = \int_0^{\Delta t} dt \cdot F_n = m_1 \Delta v_{1n}^{CG} = -m_2 \Delta v_{2n}^{CG} \quad (6)$$

and

$$J_t = \int_0^{\Delta t} dt \cdot F_t = m_1 \Delta v_{1t}^{CG} = -m_2 \Delta v_{2t}^{CG} \quad (7)$$

where the usual association between momentum-change and impulse implied by Newton's 2nd Law is used. We use Newton's 3rd law to relate the impulse on vehicle 1 to the impulse on vehicle 2.

We can further identify the "impulse ratio" by:

$$\mu = \frac{J_t}{J_n} = \frac{\Delta v_{1t}^{CG}}{\Delta v_{1n}^{CG}} = \frac{\Delta v_{2t}^{CG}}{\Delta v_{2n}^{CG}} \quad (8)$$

This ratio can be associated with frictional effects, which are directly proportional to the normal contact force. In principle, this ratio is not restricted to be less than 1.0, and can take on much larger values depending on the type of inter-vehicle surface interactions. Indeed, later we will solve for its maximum allowed value, which will depend on many factors.

Equation (8) above implies that with knowledge of $\Delta v_{kn}^{CG}$, and an estimate of $\mu$, we can solve for the magnitude of $\Delta \bar{v}_k^{CG}$ by:

$$|\Delta \bar{v}_k^{CG}| = \sqrt{(\Delta v_{kn}^{CG})^2 + (\Delta v_{kt}^{CG})^2}$$
$$= \Delta v_{kn}^{CG} \cdot \sqrt{1 + \mu^2}$$

**Restitution**

With the normal axis defined, we can now define the coefficient-of-restitution. We take this to be the negative normal projected ratio of final to initial relative velocities at the point-of-contact, given by:

$$\varepsilon = -\frac{v_{Reln,f}^{PC}}{v_{Reln,i}^{PC}} \quad (9)$$

We consider an alternative definition in the Appendix.

Again we use the subscript convention

$$v_{Reln}^{PC} = \bar{v}_{Rel}^{PC} \cdot \hat{n} \quad (10)$$

and

$$v_{Relt}^{PC} = \bar{v}_{Rel}^{PC} \cdot \hat{t} \quad (11)$$

In the limit $\varepsilon \to 0$, we obtain $v_{Reln,f}^{PC} = 0$. That is, at point $PC$ the two vehicles have reached a common velocity along the normal axis. In the limit $\varepsilon \to 1$, we obtain $v_{Reln,f}^{PC} = -v_{Reln,i}^{PC}$. In this case separation-velocity at $PC$ is equal in magnitude but opposite in direction to the closing-velocity estimated at PC along the normal axis.

With the above definition of $\varepsilon$, we can write an expression for the normal projection change-in-relative-velocity at point PC as:

$$\Delta v_{Reln}^{PC} = v_{Reln,f}^{PC} - v_{Reln,i}^{PC} \quad (12)$$

$$= -\varepsilon \cdot v_{Reln,i}^{PC} - v_{Reln,i}^{PC}$$

or

$$\boldsymbol{\Delta v_{Reln}^{PC} = -(1+\varepsilon) \cdot v_{Reln,i}^{PC}} \quad (13)$$

**Torque**

Now that we have an expression for the normal projection change-in-relative-velocity at the point-of-contact, we will need a way to ultimately arrive at the change-in-velocity at the center-of-gravity. First, we write a new expression for $\Delta v_{Reln}^{PC}$ using our earlier definition of $\bar{v}_k^P$ given by equation (1), as well as using equations (2) and (3). Here we have:

$$\bar{v}_{Rel}^{PC} = (\bar{v}_1^{CG} + \bar{\omega}_1 \times \bar{r}_1) - (\bar{v}_2^{CG} + \bar{\omega}_2 \times \bar{r}_2)$$

which can be rewritten as:

$$\bar{v}_{Rel}^{PC} = \bar{v}_{Rel}^{CG} + (\bar{\omega}_1 \times \bar{r}_1 - \bar{\omega}_2 \times \bar{r}_2) \quad (14)$$

where we have defined the closing-velocity in the usual way:

$$\bar{v}_{Rel}^{CG} = \bar{v}_1^{CG} - \bar{v}_2^{CG}$$

Now we can write an expression for the change-in-relative-velocity at the point-of-contact by:

$$\Delta \bar{v}_{Rel}^{PC} = [\bar{v}_{Rel}^{CG} + (\bar{\omega}_1 \times \bar{r}_1 - \bar{\omega}_2 \times \bar{r}_2)]_f$$
$$- [\bar{v}_{Rel}^{CG} + (\bar{\omega}_1 \times \bar{r}_1 - \bar{\omega}_2 \times \bar{r}_2)]_i$$

which can be rewritten as:

$$\Delta \bar{v}_{Rel}^{PC} = (\bar{v}_{Rel,f}^{CG} - \bar{v}_{Rel,i}^{CG}) + (\bar{\omega}_{1,f} - \bar{\omega}_{1,i}) \times \bar{r}_1$$
$$- (\bar{\omega}_{2,f} - \bar{\omega}_{2,i}) \times \bar{r}_2$$

or

$$\boldsymbol{\Delta \bar{v}_{Rel}^{PC} = \Delta \bar{v}_{Rel}^{CG} + \Delta \bar{\omega}_1 \times \bar{r}_1 - \Delta \bar{\omega}_2 \times \bar{r}_2} \quad (15)$$

where $\Delta \bar{\omega}_k$ is the change-in-angular-velocity due to torque acting on vehicle $k$ during impact.

We will solve for $\Delta v_{Reln}^{PC}$ below since this is needed for equation (13). First we must obtain useful expressions for $\Delta \bar{\omega}_k \times \bar{r}_k$.

The toque on vehicle 1 is defined by:

$$\bar{\Gamma}_1 = I_1 \bar{\alpha}_1 = \bar{r}_1 \times \bar{F} \quad (16)$$

and on vehicle 2 we have:

$$\bar{\Gamma}_2 = I_2 \bar{\alpha}_2 = \bar{r}_2 \times (-\bar{F}) \quad (17)$$

where $\bar{\alpha}_k = d\bar{\omega}_k/dt$ is the angular acceleration about the center-of-gravity of vehicle $k$, and $\bar{r}_k$ is the lever-arm extending from the center-of-gravity to the point of contact.

Taking the time integral of both sides of equations (16) and (17), we have:

$$I_1 \Delta \bar{\omega}_1 = \int_0^{\Delta t} dt \cdot \bar{\Gamma}_1 = \bar{r}_1 \times \int_0^{\Delta t} dt \cdot \bar{F}$$
$$= m_1 \bar{r}_1 \times \Delta \bar{v}_1^{CG}$$

and

$$I_2 \Delta \bar{\omega}_2 = \int_0^{\Delta t} dt \cdot \bar{\Gamma}_2 = -\bar{r}_2 \times \int_0^{\Delta t} dt \cdot \bar{F}$$
$$= -m_1 \bar{r}_2 \times \Delta \bar{v}_1^{CG}$$

Finally, solving for the changes-in-angular-velocities, we have:

$$\Delta \bar{\omega}_1 = \left(\frac{m_1}{I_1}\right) \cdot \bar{r}_1 \times \Delta \bar{v}_1^{CG} \quad (18)$$

and

$$\Delta \bar{\omega}_2 = -\left(\frac{m_1}{I_2}\right) \cdot \bar{r}_2 \times \Delta \bar{v}_1^{CG} \quad (19)$$

Using the vector triple product, we have:

$$\Delta \bar{\omega}_1 \times \bar{r}_1 = \left(\frac{m_1}{I_1}\right) \cdot \bar{r}_1 \times \Delta \bar{v}_1^{CG} \times \bar{r}_1$$
$$= \left(\frac{m_1}{I_1}\right) \cdot [\Delta \bar{v}_1^{CG}(\bar{r}_1 \cdot \bar{r}_1) - \bar{r}_1(\bar{r}_1 \cdot \Delta \bar{v}_1^{CG})] \quad (22)$$

Decomposing $\bar{r}_1$ onto the $\hat{n}$ and $\hat{t}$ axes, we have (see Figure 1):

$$\bar{r}_1 = r_{1n}\hat{n} + r_{1t}\hat{t}$$

Using equation (8) we can decompose $\Delta \bar{v}_1^{CG}$ by:

$$\Delta \bar{v}_1^{CG} = \Delta v_{1n}^{CG}\hat{n} + \Delta v_{1t}^{CG}\hat{t}$$
$$= \Delta v_{1n}^{CG} \cdot (\hat{n} + \mu\hat{t}) \quad (21)$$

The square of the magnitude of $\bar{r}_1$ is given by the dot-product:

$$\bar{r}_1 \cdot \bar{r}_1 = r_{1n}^2 + r_{1t}^2$$

We can also write the dot-product between $\bar{r}_1$ and $\Delta \bar{v}_1^{CG}$ by:

$$\bar{r}_1 \cdot \Delta \bar{v}_1^{CG} = \Delta v_{1n}^{CG} \cdot (r_{1n} + \mu r_{1t}) \quad (22)$$

Finally, taking the normal projection of $\Delta \bar{\omega}_1 \times \bar{r}_1$, we have from equation (22):

$$(\Delta \bar{\omega}_1 \times \bar{r}_1) \cdot \hat{n}$$
$$= \left(\frac{m_1}{I_1}\right) \cdot [\Delta v_{1n}^{CG}(\bar{r}_1 \cdot \bar{r}_1) - r_{1n}(\bar{r}_1 \cdot \Delta \bar{v}_1^{CG})]$$
$$= \left(\frac{m_1}{I_1}\right) \cdot [\Delta v_{1n}^{CG}(r_{1n}^2 + r_{1t}^2) - r_{1n}(\Delta v_{1n}^{CG}(r_{1n} + \mu r_{1t}))]$$
$$= \left(\frac{m_1}{I_1}\right) \cdot \Delta v_{1n}^{CG}[r_{1t}^2 - \mu r_{1t} r_{1n}] \quad (23a)$$

The tangent projection gives:

$$(\Delta \bar{\omega}_1 \times \bar{r}_1) \cdot \hat{t}$$
$$= \left(\frac{m_1}{I_1}\right) \cdot [\Delta v_{1t}^{CG}(\bar{r}_1 \cdot \bar{r}_1) - r_{1t}(\bar{r}_1 \cdot \Delta \bar{v}_1^{CG})]$$
$$= \left(\frac{m_1}{I_1}\right) \cdot [\mu \Delta v_{1n}^{CG}(r_{1n}^2 + r_{1t}^2) - r_{1t}(\Delta v_{1n}^{CG}(r_{1n} + \mu r_{1t}))]$$
$$= \left(\frac{m_1}{I_1}\right) \cdot \Delta v_{1n}^{CG}[\mu r_{1n}^2 - r_{1t} r_{1n}] \quad (23b)$$

Similarly for vehicle 2 we have:

$$\Delta \bar{\omega}_2 \times \bar{r}_2 = -\left(\frac{m_1}{I_2}\right) \cdot \bar{r}_2 \times \Delta \bar{v}_1^{CG} \times \bar{r}_2$$
$$= -\left(\frac{m_1}{I_2}\right) \cdot [\Delta \bar{v}_1^{CG}(\bar{r}_2 \cdot \bar{r}_2) - \bar{r}_2(\bar{r}_2 \cdot \Delta \bar{v}_1^{CG})]$$

and

$$(\Delta \bar{\omega}_2 \times \bar{r}_2) \cdot \hat{n}$$
$$= -\left(\frac{m_1}{I_2}\right) \cdot \Delta v_{1n}^{CG}[r_{2t}^2 - \mu r_{2t} r_{2n}] \quad (24a)$$

$$(\Delta \bar{\omega}_2 \times \bar{r}_2) \cdot \hat{t}$$
$$= -\left(\frac{m_1}{I_2}\right) \cdot \Delta v_{1n}^{CG}[\mu r_{2n}^2 - r_{2t} r_{2n}] \quad (24b)$$

Returning now to equation (15), using equations (23a) and (24a), we can rewrite the normal projected differences between angular velocity terms as:

$$(\Delta \bar{\omega}_1 \times \bar{r}_1 - \Delta \bar{\omega}_2 \times \bar{r}_2) \cdot \hat{n}$$
$$= \left(\frac{m_1}{I_1}\right) \cdot \Delta v_{1n}^{CG}[r_{1t}^2 - \mu r_{1t} r_{1n}]$$
$$+ \left(\frac{m_1}{I_2}\right) \cdot \Delta v_{1n}^{CG}[r_{2t}^2 - \mu r_{2t} r_{2n}]$$

or

$$(\Delta \bar{\omega}_1 \times \bar{r}_1 - \Delta \bar{\omega}_2 \times \bar{r}_2) \cdot \hat{n}$$
$$= m_1 \Delta v_{1n}^{CG} \cdot \left\{\left(\frac{r_{1t}^2}{I_1} + \frac{r_{2t}^2}{I_2}\right) - \mu\left(\frac{r_{1t} r_{1n}}{I_1} + \frac{r_{2t} r_{2n}}{I_2}\right)\right\}$$

Now using (15) and rewriting the expression for $(\Delta \bar{v}_{Rel}^{PC} \cdot \hat{n})$, we finally have:

$$\Delta v_{Reln}^{PC} = \Delta v_{Reln}^{CG}$$
$$+ m_1 \Delta v_{1n}^{CG}$$
$$\times \left\{\left(\frac{r_{1t}^2}{I_1} + \frac{r_{2t}^2}{I_2}\right) - \mu\left(\frac{r_{1t} r_{1n}}{I_1} + \frac{r_{2t} r_{2n}}{I_2}\right)\right\} \quad (25)$$

With this our work is nearly finished. First, we will boost to the center-of-mass frame of reference in order to simplify our equations.

**The Center-of-Mass Frame**

We identify the center-of-mass of the two-vehicle system by:

$$\bar{R}_{CM} = \frac{m_1 \bar{R}_1 + m_2 \bar{R}_2}{m_1 + m_2}$$

where $\bar{R}_k$ is the position vector extending from the Earth-frame's origin to the center-of-gravity of vehicle $k$. Taking the time derivative of both sides of this equation, we have for any time $t$:

$$\bar{V}_{CM}(t) = \frac{m_1 \bar{v}_1^{CG}(t) + m_2 \bar{v}_2^{CG}(t)}{m_1 + m_2} \qquad (26)$$

Without any loss of generality, we can switch to the center-of-mass frame by making the transformation for vehicle 1:

$$\bar{v}_1^{CG}(t) \to \bar{v}_1^{CG'}(t) = \bar{v}_1^{CG}(t) - \bar{V}_{CM}(t)$$
$$= \left(\frac{m_2}{m_1 + m_2}\right) \cdot \left(\bar{v}_1^{CG}(t) - \bar{v}_2^{CG}(t)\right)$$
$$= \left(\frac{\bar{m}}{m_1}\right) \cdot \bar{v}_{Rel}^{CG}(t) \qquad (27)$$

Where we identify the reduced mass of the two-vehicle system by

$$\bar{m} = \frac{m_1 m_2}{m_1 + m_2}$$

Similarly, for vehicle 2, we have:

$$\bar{v}_2^{CG}(t) \to \bar{v}_2^{CG'}(t) = \bar{v}_2^{CG}(t) - \bar{V}_{CM}(t)$$
$$= -\left(\frac{m_1}{m_1 + m_2}\right) \cdot \left(\bar{v}_1^{CG}(t) - \bar{v}_2^{CG}(t)\right)$$
$$= -\left(\frac{\bar{m}}{m_2}\right) \cdot \bar{v}_{Rel}^{CG}(t) \qquad (28)$$

Here we use the prime symbol to denote that the quantity is estimated in the center-of-mass frame. Conservation of linear momentum is guaranteed as we have:

$$m_1 \bar{v}_1^{CG'}(t) + m_2 \bar{v}_2^{CG'}(t)$$
$$= m_1 \left[\left(\frac{\bar{m}}{m_1}\right) \cdot \bar{v}_{Rel}^{CG}(t)\right] + m_2 \left[-\left(\frac{\bar{m}}{m_2}\right) \cdot \bar{v}_{Rel}^{CG}(t)\right] = 0$$

Therefore, this expression is always 0 for any time $t$. Since Newton's laws are *Galilean invariant*, the collision force and therefore delivered impulse remain unchanged in the center-of-mass frame:

$$\bar{J} = \int_0^{\Delta t} dt \cdot \bar{F} = \int_0^{\Delta t} dt \cdot \bar{F}' = \bar{J}'$$

Since impulse is equal to momentum change, using (27), for vehicle 1 we have:

$$\bar{J} = m_1 \Delta \bar{v}_1^{CG} = m_1 \Delta \bar{v}_1^{CG'}$$
$$= m_1 \left(\frac{\bar{m}}{m_1}\right) \cdot \Delta \bar{v}_{Rel}^{CG} = \bar{m} \Delta \bar{v}_{Rel}^{CG}$$

For vehicle 2 we have:

$$-\bar{J} = m_2 \Delta \bar{v}_2^{CG} = m_2 \Delta \bar{v}_2^{CG'}$$
$$= -m_2 \left(\frac{\bar{m}}{m_2}\right) \cdot \Delta \bar{v}_{Rel}^{CG} = -\bar{m} \Delta \bar{v}_{Rel}^{CG}$$

Therefore, we have the very useful relations which will allow us to transform from center-of-mass frame to Earth-frame:

$$\Delta \bar{v}_1^{CG} = \left(\frac{\bar{m}}{m_1}\right) \Delta \bar{v}_{Rel}^{CG}$$
$$\Delta \bar{v}_2^{CG} = -\left(\frac{\bar{m}}{m_2}\right) \Delta \bar{v}_{Rel}^{CG} \qquad (29)$$

With these, we can now return to simplifying equation (25). Making the substitution for $\Delta \bar{v}_1^{CG}$ given by (29) and taking the normal projection, we have:

$$\Delta v_{Reln}^{PC} = \Delta v_{Reln}^{CG} + m_1 \left[\left(\frac{\bar{m}}{m_1}\right) \Delta v_{Reln}^{CG}\right]$$
$$\times \left\{\left(\frac{r_{1t}^2}{I_1} + \frac{r_{2t}^2}{I_1}\right) - \mu \left(\frac{r_{1t} r_{1n}}{I_1} + \frac{r_{2t} r_{2n}}{I_1}\right)\right\}$$

Factoring out $\Delta \bar{v}_{Reln}^{CG}$ and rewriting this expression, we have the key result which allows us to relate the change-in-velocity estimated at *PC* to that estimated at *CG*:

$$\Delta v_{Reln}^{PC} = \left(\frac{1}{q}\right) \cdot \Delta v_{Reln}^{CG} \qquad (30)$$

where we have adopted Brach's definitions of the dimensionless rotation parameters:

$$\left(\frac{1}{q}\right) = A - \mu B$$

$$A = 1 + \bar{m} \left(\frac{r_{1t}^2}{I_1} + \frac{r_{2t}^2}{I_2}\right)$$

$$B = \bar{m} \left(\frac{r_{1t} r_{1n}}{I_1} + \frac{r_{2t} r_{2n}}{I_2}\right) \qquad (31)$$

**Normal Change-in-Velocity**

Using equations (13) and (30), we can finally write an expression for the normal projection *CG* change-in-velocity in terms of the normal projection closing-speed at *PC*:

$$\Delta v_{Reln}^{PC} = -(1 + \varepsilon) \cdot v_{Reln,i}^{PC}$$
$$= \Delta v_{Reln}^{CG}/q$$

Therefore we have:

$$\Delta v_{Reln}^{CG} = -(1 + \varepsilon) \cdot q \cdot v_{Reln,i}^{PC} \qquad (32)$$

We can relate this back to the changes-in-velocity for vehicles 1 and 2 by using equation (29):

$$\Delta v_{1n}^{CG} = -\left(\frac{\bar{m}}{m_1}\right) \cdot (1 + \varepsilon) \cdot q \cdot v_{Reln,i}^{PC}$$

$$\Delta v_{2n}^{CG} = \left(\frac{\bar{m}}{m_2}\right) \cdot (1 + \varepsilon) \cdot q \cdot v_{Reln,i}^{PC} \qquad (33)$$

For most particle applications, equations (33) are of crucial importance as they allow one to estimate the *CG* change-in-velocity with knowledge of the closing-speed at impact and the geometrical properties of the impact orientation. To estimate the magnitude of the *total* change-in-velocity $\Delta \bar{v}_k^{CG}$ however, one must use the critical impulse ratio in equation (8), whose magnitude is bound from above due to the common velocity assumption. This upper bound is derived next.

**The Critical Impulse Ratio**

Taking the tangent projection of $\Delta \bar{v}_{Rel}^{PC}$, we have from (15):

$$\Delta v_{Relt}^{PC} = \Delta v_{Relt}^{CG} + (\Delta \bar{\omega}_1 \times \bar{r}_1 - \Delta \bar{\omega}_2 \times \bar{r}_2) \cdot \hat{t}$$

Using (8), (23b), (24b), (29), and (31), we can write this as:

$$\Delta v_{Relt}^{PC} = \mu \Delta v_{Reln}^{CG}$$
$$+ \left(\frac{m_1}{I_1}\right) \cdot \Delta v_{1n}^{CG} [\mu r_{1n}^2 - r_{1t} r_{1n}]$$
$$+ \left(\frac{m_1}{I_2}\right) \cdot \Delta v_{1n}^{CG} [\mu r_{2n}^2 - r_{2t} r_{2n}]$$
$$= \mu \Delta v_{Reln}^{CG} + m_1 \Delta v_{1n}^{CG}$$
$$\times \left\{\mu \left(\frac{r_{1n}^2}{I_1} + \frac{r_{2n}^2}{I_2}\right) - \left(\frac{r_{1t} r_{1n}}{I_1} + \frac{r_{2t} r_{2n}}{I_2}\right)\right\}$$
$$= \Delta v_{Reln}^{CG} \cdot \{\mu(1 + C) - B\} \qquad (34a)$$

where Brach's dimensionless *C* coefficient is defined by:

$$C = \bar{m} \left(\frac{r_{1n}^2}{I_1} + \frac{r_{2n}^2}{I_2}\right)$$

In particular, we are interested in the case where the tangent projection relative motion at the point of contact goes to 0. That is, the case where the relative sliding motion tangent to the surface-of-contact is fully retarded by frictional effects. That is, in the limit: $v_{Relt,f}^{PC} \to 0$. In this case, using (32) and our prior expression we have:

$$\Delta v_{Relt}^{PC} = -v_{Relt,i}^{PC} = -r \cdot v_{Reln,i}^{PC}$$
$$= \Delta v_{Reln}^{CG} \cdot \{\mu(1 + C) - B\}$$
$$= [-(1 + \varepsilon) \cdot q \cdot v_{Reln,i}^{PC}] \cdot \{\mu(1 + C) - B\}$$

where we also use Brach's dimensionless parameter *r*:

$$r = \frac{v_{Relt,i}^{PC}}{v_{Reln,i}^{PC}}$$

Dividing both sides by $v_{Reln,i}^{PC}$ and isolating $\mu$, we get:

$$\mu = \frac{r}{(1 + \varepsilon)(1 + C)q} + \frac{B}{1 + C}$$

Using (31) to substitute for $(1/q)$ and multiplying the second term by $\frac{1+\varepsilon}{1+\varepsilon}$, this gives:

$$\mu = \frac{r(A - \mu B) + B(1 + \varepsilon)}{(1 + \varepsilon)(1 + C)}$$

Factoring out $\mu$, we have:

$$\mu \cdot \left(1 + \frac{rB}{(1 + \varepsilon)(1 + C)}\right) = \frac{rA + B(1 + \varepsilon)}{(1 + \varepsilon)(1 + C)}$$

Solving for $\mu$:

$$\mu = \left(\frac{(1 + \varepsilon)(1 + C)}{rB + (1 + \varepsilon)(1 + C)}\right) \times \frac{rA + B(1 + \varepsilon)}{(1 + \varepsilon)(1 + C)}$$

Finally, we have the critical impulse ratio:

$$\boldsymbol{\mu_{max}} = \frac{\boldsymbol{rA + B(1 + \varepsilon)}}{\boldsymbol{rB + (1 + \varepsilon)(1 + C)}} \qquad (34b)$$

This value represents the impulse ratio which maximizes energy losses due to tangential effects. Therefore, one will typically define the impulse ratio to represent some fraction of the maximum value derived from equation (34b). At the maximum value, all relative tangential motion will stop, and the vehicles will reach common tangential velocity at *PC*.

**Change-in-Velocity from Crush Energy Approximation**

Suppose, as is generally the case, our subject vehicles exhibit crush damage, and we wish to use this damage to estimate the closing-speed of impact and changes-in-velocity. Let us also suppose that the normal axis, $\hat{n}$, is anti-parallel with the surface normal vector at point *PC* on vehicle 1 (see Figures 1 and 2). We can estimate the differential work done to impart crush to both vehicles using the expression:

$$dW_n = -dE_n = dn_1 \cdot F_{n1} + dn_2 \cdot F_{n2}$$

In the tangent direction, we have:

$$dW_t = -dE_t = dt_1 \cdot F_{t1} + dt_2 \cdot F_{t2}$$

Where $n_1$ and $n_2$ are the normal projection deflection vectors characterizing crush damage on vehicles 1 and 2 respectively, and $t_1$ and $t_2$ are the distances traveled along the tangent direction. Using Newton's 3rd law, we have

$$F_n = F_{n1} = -F_{n2}$$

This implies:

$$-dE_n = (dn_1 - dn_2) \cdot F_n$$

Let $N = n_1 - n_2$, which is a measure of the total crush imparted across both vehicles. Here it is understood that $n_1$ and $n_2$ have opposite signs. With this definition of $N$, this implies $dN = dn_1 - dn_2$. Using this we have:

$$-dE_n = dN \cdot F_n$$
$$= m_1 dN \cdot \frac{dv_{1n}^{CG}}{dt} = m_1 dv_{1n}^{CG} \cdot \frac{dN}{dt}$$
$$= m_1 dv_{1n}^{CG} \cdot v_{Reln}^{PC}$$

where we identify the time-rate-of-change of the total normal crush with the normal projection relative-velocity: $\frac{dN}{dt} = v_{Reln}^{PC}$. Integrating both sides, we have:

$$-E_n = \int m_1 dv_{1n}^{CG} \cdot v_{Reln}^{PC}$$
$$\approx \langle v_{Reln}^{PC} \rangle \cdot \int m_1 dv_{1n}^{CG}$$

Where $\langle v_{Reln}^{PC} \rangle$ is the normal projection average relative velocity near the point-of-contact. The energy lost can then be approximated by:

$$E_n \approx -m_1 \langle v_{Reln}^{PC} \rangle \cdot \Delta v_{1n}^{CG}$$

Using this approximation allows the planar impact dynamics equations to become tractable and much easier to manipulate. The approximation also allows us to easily "factorize" effects related to restorative forces versus those that are purely dissipative in the tangent direction.

Using the normal projection of equation (29), this becomes:

$$E_n \approx -\bar{m} \langle v_{Reln}^{PC} \rangle \cdot \Delta v_{Reln}^{CG} \qquad (35a)$$

Similarly, along the tangent direction, we have:

$$E_t \approx -\bar{m} \langle v_{Relt}^{PC} \rangle \cdot \Delta v_{Relt}^{CG} \qquad (35b)$$

Writing an expression for $\langle v_{Reln}^{PC} \rangle$, we have:

$$\langle v_{Reln}^{PC} \rangle = \frac{1}{2} \left( v_{Reln,i}^{PC} + v_{Reln,f}^{PC} \right) \qquad (36a)$$

and for $\langle v_{Relt}^{PC} \rangle$, we have:

$$\langle v_{Relt}^{PC} \rangle = \frac{1}{2} \left( v_{Relt,i}^{PC} + v_{Relt,f}^{PC} \right)$$
$$= \frac{1}{2} \left( 2 v_{Relt,i}^{PC} + \Delta v_{Relt}^{PC} \right) \qquad (36b)$$

Using equation (9), (36a) becomes:

$$\langle v_{Reln}^{PC} \rangle = \frac{1}{2} v_{Reln,i}^{PC} (1 - \varepsilon) \qquad (37a)$$

Using equations (8), (32), (34a), and our expression for $r$, equation (36b) becomes:

$$\langle v_{Relt}^{PC} \rangle = r \cdot v_{Reln,i}^{PC} + \frac{1}{2} \left[ -(1+\varepsilon) \cdot q \cdot v_{Reln,i}^{PC} \right]$$
$$\times \{\mu(1+C) - B\}$$
$$= v_{Reln,i}^{PC} \times \left\{ r - \frac{1}{2}(1+\varepsilon) q [\mu(1+C) - B] \right\} \qquad (37b)$$

We can solve for the energy losses associated with both normal and tangential effects.

First, using (32), (35a), and (37a). We have for the energy losses from normal effects:

$$E_n(\varepsilon) \approx -\bar{m} \cdot \left[ \frac{1}{2} v_{Reln,i}^{PC} (1-\varepsilon) \right]$$
$$\times \left[ -(1+\varepsilon) \cdot q \cdot v_{Reln,i}^{PC} \right]$$
$$= \frac{1}{2} \bar{m} \left( v_{Reln,i}^{PC} \right)^2 \cdot q \cdot (1 - \varepsilon^2) \qquad (38a)$$

Similarly, energy losses related to tangent effects can be written by:

$$E_t \approx -\bar{m} v_{Reln,i}^{PC} \times \left\{ r - \frac{1}{2}(1+\varepsilon) q [\mu(1+C) - B] \right\}$$
$$\times \mu \left\{ -(1+\varepsilon) \cdot q \cdot v_{Reln,i}^{PC} \right\}$$
$$= \frac{1}{2} \bar{m} \left( v_{Reln,i}^{PC} \right)^2 \cdot q \cdot (1+\varepsilon)$$
$$\times \{2\mu r - \mu(1+\varepsilon) q [\mu(1+C) - B]\}$$

Summing both the normal and tangent contributions, we arrive at an expression for the total energy lost:

$$\boldsymbol{E_{Loss}(\varepsilon, \mu) \approx \frac{1}{2} \bar{m} \left( v_{Reln,i}^{PC} \right)^2 \cdot q \cdot (1+\varepsilon)}$$
$$\boldsymbol{\times \{1 - \varepsilon + 2\mu r - \mu(1+\varepsilon) q [\mu(1+C) - B]\}}$$

This expression will be examined later to one given by the exact solution to the equations of motion.

For now, returning to the $E_n(\varepsilon)$, we note it is at its maximum, $E_n^{Max}$, when $\varepsilon = 0$. We can then write an expression for $E_n^{Max}$:

$$\boldsymbol{E_n^{Max} \approx \frac{1}{2} q \cdot \bar{m} \left( v_{Reln,i}^{PC} \right)^2} \qquad (38b)$$

This can be interpreted as the energy absorbed by the normal effects at the moment of maximum engagement. Assuming we can estimate $E_n^{Max}$ from residual crush on the subject vehicles, we can work backward and estimate the normal projection closing-speed at the point-of-contact:

$$v_{Reln,i}^{PC} \approx \sqrt{\frac{2 E_n^{Max}}{\bar{m} q}} \qquad (39a)$$

In the special case where the impulse is delivered through the centers-of-gravity of both vehicles, we have in the zero-net-torque limit:

$$\left( \frac{1}{q} \right) \to 1$$

In this case, we have:

$$v_{Reln,i}^{PC} \to v_{Reln,i}^{PC*} \approx \sqrt{\frac{2 E_n^{Max}}{\bar{m}}} \qquad (39b)$$

With this, we can rewrite (39a):

$$v_{Reln,i}^{PC} \approx \frac{v_{Reln,i}^{PC*}}{\sqrt{q}} \qquad (40)$$

Here we interpret $v_{Reln,i}^{PC*}$ as the equivalent normal projection closing-speed needed to produce the identical amount of damage as in our given subject case, but with no net torque.

Finally, returning to (32), and using (40), we have:

$$\Delta v_{Reln}^{CG} \approx -(1+\varepsilon) \cdot \sqrt{q} \cdot v_{Reln,i}^{PC*} \qquad (41)$$

Now that we have a convenient way to solve for the change-in-velocity given an estimate of the equivalent closing-speed, we move on to using crush damage to estimate the closing-speed.

**Closing-Speed from Crush Damage**

Here we re-derive Campbell's standard method for estimating energy dissipated by crush damage. First, let us assume we can model the force-deflection response of both vehicles as an array of discrete linear springs pointing along the normal direction, and satisfying Hooke's law. Along the surface of contact of vehicle 1, for any spring *j*, the magnitude of the forces will satisfy:

$$F_{1n,j} = k_{1j} \cdot C_{1,j} \qquad (43)$$

and for vehicle 2 we have:

$$F_{2n,j} = k_{2j} \cdot C_{2,j} = F_{1n,j}$$

Here $C_{1,j}$ and $C_{2,j}$ are measures of the amount of permanent deflection in the $j^{th}$ spring, and $k_{1j}$ and $k_{2j}$ are the linear spring constants. This allows us to solve for the crush on vehicle 2 in terms of the crush on vehicle 1:

$$C_{2,j} = \frac{k_{1j}}{k_{2j}} C_{1,j} \qquad (44)$$

The energy absorbed by the $j^{th}$ spring of vehicles 1 and 2 is given by:

$$E_j = \frac{1}{2} k_{1j} C_{1,j}^2 + \frac{1}{2} k_{2j} C_{2,j}^2$$
$$= \frac{1}{2} k_{1j} C_{1,j}^2 + \frac{1}{2} k_{2j} \left(\frac{k_{1j}}{k_{2j}} C_{1,j}\right)^2$$
$$= \frac{1}{2} \left(\frac{k_{1j} + k_{2j}}{k_{1j} k_{2j}}\right) \cdot k_{1j}^2 \cdot C_{1,j}^2 \qquad (45)$$

The total energy absorbed is given by the sum:

$$E = \sum_j \frac{1}{2} \left(\frac{k_{1j} + k_{2j}}{k_{1j} k_{2j}}\right) \cdot k_{1j}^2 \cdot C_{1,j}^2 \qquad (46)$$

In the infinitesimal limit, we have:

$$E \to \int dw \cdot \frac{1}{2} \left(\frac{B_1 + B_2}{B_1 B_2}\right) \cdot B_1^2 \cdot C_1^{Max}(w)^2 \qquad (47)$$

where $k \to B$ is the effective vehicle stiffness per unit of width and $w$ specifies the position within the crush profile along which the crush damage is being measured. $C_1^{Max}$ is the maximum crush imparted to vehicle 1 at some position $w$ (Figure 3). We can rewrite $C_1^{Max}$ in terms of the measured permanent crush using:

$$C_1^{Max}(w) = \frac{A_1}{B_1} + C_1(w) \qquad (48)$$

Here $A_1$ is the standard "A" coefficient, typically interpreted as the pre-load force per unit width needed to cause permanent crush beyond the elastic limit. Similarly, the ratio $\frac{A_1}{B_1}$ is generally taken as elastic limit of deformation beyond which there is permanent crush. $C_1(w)$ is the measured amount of permanent crush at point $w$ along the damage profile, with $C_1(w)$ being orthogonal to $w$ (Figure 3). Combining (47) and (48) our expression for energy absorbed at maximum engagement is:

$$E = \int dw \cdot \frac{1}{2} \left(\frac{B_1 + B_2}{B_1 B_2}\right) \cdot B_1^2 \cdot \left(\frac{A_1}{B_1} + C_1(w)\right)^2 \qquad (49)$$

As usual, without loss of generality, we can write the same expression for total energy absorbed, estimated by the crush on vehicle 2, by switching the indices 1↔2.

Let us assume that we can measure the permanent crush imparted to vehicle 1 such that our crush measurements are taken in the standard way, parallel to the vehicle contact surface normal vector. Let us define our normal axis, $\hat{n}$, to run anti-parallel with this surface normal. We can then relate equation (39b) to (49) by assuming the total absorbed energy is equal to the term $E_n^{Max}$. With this we can solve for $v_{Reln,i}^{PC*}$ by:

$$v_{Reln,i}^{PC*} = \left\{ \frac{2}{\bar{m}} \int dw \cdot \frac{1}{2} \left(\frac{B_1 + B_2}{B_1 B_2}\right) \cdot B_1^2 \cdot \left(\frac{A_1}{B_1} + C_1(w)\right)^2 \right\}^{1/2}$$

With respect to damage imparted to vehicle 2, this becomes:

$$v_{Reln,i}^{PC*} = \left\{ \frac{2}{\bar{m}} \int dw \cdot \frac{1}{2} \left(\frac{B_1 + B_2}{B_1 B_2}\right) \cdot B_2^2 \cdot \left(\frac{A_2}{B_2} + C_2(w)\right)^2 \right\}^{1/2}$$

With these final relations, we can fully characterize the dynamics of a vehicle impact given crush damage on one vehicle or the other. Next we given an example application.

**An Example Impact**

Let us return to the scenario depicted in Figure 1. Here a 2013 Porche 911 Carrera impacts a 2008 Bentley Arnage near the rear axle of the Bentley. Using a graphical tool such as ARAS 360 HD [5], we easily can estimate $PC$ in the $n$-$t$ coordinate system for this impact. Here we have the following inputs for our calculation:

Vehicle 1: 2008 Bentley Arnage 4 door
$W_1 = 5750 \; lb$ (curb weight)
$I_1 = 4716.5 \; lb \cdot ft \cdot s^2$ (yaw moment-of-inertia)
$r_{1n} = 3.1 \; ft$
$r_{1t} = 10.2 \; ft$
$V_{1n} = 0 \; mph$
$V_{1t} = -10 \; mph$

Vehicle 2: 2013 Porsche 911 Carrera 2 door
$W_2 = 3100 \; lb$ (curb weight)
$I_2 = 1987 \; lb \cdot ft \cdot s^2$ (yaw moment-of-inertia)
$r_{2n} = -5.8 \; ft$
$r_{2t} = -4.7 \; ft$
$V_{2n} = -21.2 \; mph$
$V_{1t} = -21.2 \; mph$

Note here we are using the curb weights and moments-of-Inertias for both vehicles as reported by Expert AutoStats [6].

Solving for the reduced-mass, we have:

$$\bar{m} = \frac{m_1 m_2}{m_1 + m_2} = 62.6 \; slug$$

From equation (2) we have:

$$\bar{v}_{Rel,i}^{PC} = \bar{v}_{1,i}^{PC} - \bar{v}_{2,i}^{PC} = (0 \; mph + 21.2 \; mph)\hat{n}$$
$$+(-10 \; mph + 21.2 \; mph)\hat{t}$$
$$= (21.2 \; mph)\hat{n} + (11.2 \; mph)\hat{t}$$

The ratio of tangent-to-normal initial relative velocities is given by:

$$r = \frac{v_{Relt,i}^{PC}}{v_{Reln,i}^{PC}} = \frac{11.2}{21.2} = 0.528$$

Here we note $\tan^{-1}(r) \approx 27.8°$. This is the angle between the $\hat{n}$ axis and the vector $\bar{v}_{Rel,i}^{PC}$. That is, the relative velocity vector is 27.8° away from being perfectly aligned with the $\hat{n}$ axis.

Now we can solve for our spin terms defined by (31):

$$A = 1 + \bar{m}\left(\frac{r_{1t}^2}{I_1} + \frac{r_{2t}^2}{I_2}\right)$$
$$= 1 + 62.6 \; slug$$
$$\times \left[\frac{(10.2 \; ft)^2}{(4716.5 \; lb \cdot ft \cdot s^2)} + \frac{(-4.7 \; ft)^2}{(1987 \; lb \cdot ft \cdot s^2)}\right]$$
$$= 3.1$$

$$B = \bar{m}\left(\frac{r_{1t} r_{1n}}{I_1} + \frac{r_{2t} r_{2n}}{I_2}\right)$$
$$= 62.6 \; slug$$
$$\times \left[\frac{(10.2 \; ft \times 3.1 \; ft)}{(4716.5 \; lb \cdot ft \cdot s^2)} + \frac{(-4.7 \; ft \times -5.8 \; ft)}{(1987 \; lb \cdot ft \cdot s^2)}\right]$$
$$= 1.3$$

$$C = \bar{m}\left(\frac{r_{1n}^2}{I_1} + \frac{r_{2n}^2}{I_2}\right)$$
$$= 62.6 \; slug$$
$$\times \left[\frac{(3.1 \; ft)^2}{(4716.5 \; lb \cdot ft \cdot s^2)} + \frac{(-5.8 \; ft)^2}{(1987 \; lb \cdot ft \cdot s^2)}\right]$$
$$= 1.2$$

The critical impulse ratio is given by equation (34b):

$$\mu_{max} = \frac{rA + B(1+\varepsilon)}{rB + (1+\varepsilon)(1+C)}$$
$$= \frac{0.528 \times 3.1 + 1.3(1+\varepsilon)}{0.528 \times 1.3 + (1+\varepsilon)(1+1.2)}$$
$$= \frac{1.6 + 1.3 \cdot (1+\varepsilon)}{0.7 + 2.2 \cdot (1+\varepsilon)}$$

This is shown in Figure 4, where $\mu_{max}$ is plotted as a function of $\varepsilon$ for our example impact. Note, the minimum value for $\mu_{max}(\varepsilon = 1) = 0.83$. Its maximum value is given by $\mu_{max}(\varepsilon = 0) = 1.01$. It is to be expected that the maximum allowed impulse ratio goes as the inverse of restitution since tangential forces imparted during the restoration phase of collision must also be accounted for.

Figure 5 shows how the value of $q$ varies with both $\varepsilon$ and $\mu$, where $\mu$ is expressed as a fraction of the maximum allowed value, $\mu_{max}$.

Figure 6 shows the resulting final normal and tangent velocity components for vehicle 1 versus both $\varepsilon$ and $\mu$. As expected, the magnitude of the normal component increases in direct proportion to $\varepsilon$. We also note the normal projection increases in magnitude with $\mu$. Examining Figure 1 may provide some insight as to why this is the case. As the impulse ratio increases away from 0, the total force vector sweeps closer to the $CG$ of vehicle 1. One naively expects that as the force vector is directed through the $CG$, this will tend to increase the resulting change-in-velocity at the $CG$.

We also note the tangent component remains constant in the no-friction limit, but its magnitude increases with increasing value of $\mu$. This of course is sensible as increased friction during the contact will tend to boost vehicle 1 along its original direction of travel in our collision scenario.

Figure 7 shows the resulting magnitude of the $|\Delta \bar{v}_{Rel}^{CG}|$ versus both $\varepsilon$ and $\mu$. On the top we show the values for the scenario depicted in Figure 1. On the bottom we show how this result changes as we allow for variation in $r_{1t}$ by letting $r_{1t}$ be Gaussian distributed with mean value 10.17 ft and sigma = ½ ft. This type of monte carlo analysis is trivial in ROOT [7]. As one should expect, a variation in *PC* will drive variations in the geometrical parameters *A,B,C*, and *q*. This of course will cause variations in the final velocities of both vehicles. A resulting smearing of order 1 mph is observed throughout the distributions shown in Figure 7.

Suppose we focus in on the region where $\varepsilon < 0.1$, close to the common-velocity condition along the normal axis. We can now examine the frequency at which particular values of $|\Delta \bar{v}_{Rel}^{CG}|$ as we allow $r_{1t}$ and $\varepsilon$ to randomly vary. This is shown in Figure 8. These distributions could easily be fit with Gaussian curves to obtain reasonable estimates for $|\Delta \bar{v}_{Rel}^{CG}|$.

We thus finish our practical exercise in use of the derived equations. We now to turn our attention to the question of what effect if any, the use of the approximate energy loss terms have on our calculations.

**An Exact Solution to Change-in-Velocity from Energy Loss**

Now we will derive an exact solution for change-in-velocity about the center-of-gravity, where the total energy lost is assumed known. First, we start with an expression of energy conservation, relating the initial kinetic and rotational energy terms to the final kinetic and rotational energy terms, in the Earth-frame:

$$\frac{1}{2}m_1(\bar{v}_{1i}^{CG})^2 + \frac{1}{2}m_2(\bar{v}_{2i}^{CG})^2$$
$$+ \frac{1}{2}I_1(\bar{\omega}_{1i})^2 + \frac{1}{2}I_1(\bar{\omega}_{1i})^2$$
$$= \frac{1}{2}m_1(\bar{v}_{1f}^{CG})^2 + \frac{1}{2}m_2(\bar{v}_{2f}^{CG})^2$$
$$+ \frac{1}{2}I_1(\bar{\omega}_{1f})^2 + \frac{1}{2}I_1(\bar{\omega}_{1f})^2 + E_{Loss} \quad (50)$$

Using equations (27) and (28), we once again switch to the center-of-mass frame of reference, and multiply by sides by $(2/\bar{m})$. This gives:

$$(\bar{v}_{Rel,i}^{CG})^2 + \frac{I_1}{\bar{m}}(\bar{\omega}_{1i})^2 + \frac{I_2}{\bar{m}}(\bar{\omega}_{2i})^2 =$$
$$(\bar{v}_{Rel,f}^{CG})^2 + \frac{I_1}{\bar{m}}(\bar{\omega}_{1f})^2 + \frac{I_2}{\bar{m}}(\bar{\omega}_{2f})^2 + \frac{2}{\bar{m}}E_{Loss} \quad (51)$$

We can rewrite this as:

$$\left[(\bar{v}_{Rel,f}^{CG})^2 - (\bar{v}_{Rel,i}^{CG})^2\right] + \frac{I_1}{\bar{m}} \cdot \left[(\bar{\omega}_{1f})^2 - (\bar{\omega}_{1i})^2\right]$$
$$+ \frac{I_2}{\bar{m}} \cdot \left[(\bar{\omega}_{2f})^2 - (\bar{\omega}_{2i})^2\right] + \frac{2}{\bar{m}}E_{Loss} = 0 \quad (52)$$

Using simple vector algebra, we have the following relations:

$$\bar{v}_{Rel,f}^{CG} = \bar{v}_{Rel,i}^{CG} + \Delta \bar{v}_{Rel}^{CG} \quad (53)$$

and

$$\bar{\omega}_{kf} = \bar{\omega}_{ki} + \Delta \bar{\omega}_k \quad (54)$$

where *k* is the usual vehicle index.

Taking the dot-product of equation (53) with itself, we have:

$$(\bar{v}_{Rel,f}^{CG})^2 = (\bar{v}_{Rel,i}^{CG} + \Delta \bar{v}_{Rel}^{CG}) \cdot (\bar{v}_{Rel,i}^{CG} + \Delta \bar{v}_{Rel}^{CG})$$
$$= (\bar{v}_{Rel,i}^{CG})^2 + (\Delta \bar{v}_{Rel}^{CG})^2 + 2\bar{v}_{Rel,i}^{CG} \cdot \Delta \bar{v}_{Rel}^{CG}$$

or

$$(\bar{v}_{Rel,f}^{CG})^2 - (\bar{v}_{Rel,i}^{CG})^2 = (\Delta \bar{v}_{Rel}^{CG})^2 + 2\bar{v}_{Rel,i}^{CG} \cdot \Delta \bar{v}_{Rel}^{CG} \quad (55)$$

Similarly for equation (54), we have:

$$(\bar{\omega}_{kf})^2 - (\bar{\omega}_{ki})^2 = (\Delta \bar{\omega}_k)^2 + 2\bar{\omega}_{ki} \cdot \Delta \bar{\omega}_k \quad (56)$$

From equations (18), (19), and (29), we have:

$$(\Delta \bar{\omega}_k)^2 = \left(\frac{\bar{m}}{I_k}\right)^2 \cdot |\bar{r}_k \times \Delta \bar{v}_{Rel}^{CG}|^2 \quad (57)$$

We can write the above cross-product by:

$$\bar{r}_k \times \Delta \bar{v}_{Rel}^{CG} = \begin{vmatrix} \hat{n} & \hat{t} & \hat{z} \\ r_{kn} & r_{kt} & 0 \\ \Delta v_{Reln}^{CG} & \mu \Delta v_{Reln}^{CG} & 0 \end{vmatrix}$$

$$= \hat{z}[\mu r_{kn} - r_{kt}] \cdot \Delta v_{Reln}^{CG} \quad (58)$$

Giving us:

$$(\Delta \bar{\omega}_k)^2 = \left(\frac{\bar{m}}{I_k}\right)^2 \cdot [\mu r_{kn} - r_{kt}]^2 \cdot (\Delta v_{Reln}^{CG})^2 \quad (59)$$

The second term in equation (56) can be expressed by:

$$\bar{\omega}_{ki} \cdot \Delta \bar{\omega}_k = \left(\frac{\bar{m}}{I_k}\right) \bar{\omega}_{ki} \cdot (\bar{r}_k \times \Delta \bar{v}_{Rel}^{CG})$$
$$= \left(\frac{\bar{m}}{I_k}\right) \cdot \omega_{ki}[\mu r_{kn} - r_{kt}] \cdot \Delta v_{Reln}^{CG} \quad (60)$$

Returning to equation (52), using (59) and (60), we can write:

$$\frac{I_k}{\bar{m}} \cdot \left[(\bar{\omega}_{kf})^2 - (\bar{\omega}_{ki})^2\right]$$
$$= \left(\frac{\bar{m}}{I_k}\right) \cdot [\mu r_{kn} - r_{kt}]^2 \cdot (\Delta v_{Reln}^{CG})^2$$
$$+ 2\omega_{ki} \cdot [\mu r_{kn} - r_{kt}] \cdot \Delta v_{Reln}^{CG} \quad (61)$$

We can now write the change-in-rotational-energy terms as:

$$\frac{I_1}{\bar{m}} \cdot \left[(\bar{\omega}_{1f})^2 - (\bar{\omega}_{1i})^2\right] + \frac{I_2}{\bar{m}} \cdot \left[(\bar{\omega}_{2f})^2 - (\bar{\omega}_{2i})^2\right]$$
$$= (\Delta v_{Reln}^{CG})^2 \times \left\{ \begin{array}{l} \left(\frac{\bar{m}}{I_1}\right) \cdot [\mu r_{1n} - r_{1t}]^2 \\ + \left(\frac{\bar{m}}{I_2}\right) \cdot [\mu r_{2n} - r_{2t}]^2 \end{array} \right\}$$
$$+ 2\Delta v_{Reln}^{CG} \times \{\omega_{1i} \cdot [\mu r_{1n} - r_{1t}] + \omega_{2i} \cdot [\mu r_{2n} - r_{2t}]\}$$

$$= (\Delta v_{Reln}^{CG})^2 \cdot \{\mu^2 C + (A - 1) - 2\mu B\}$$
$$+ 2\Delta v_{Reln}^{CG} \times \{\omega_{1i} \cdot \tilde{d}_1 + \omega_{2i} \cdot \tilde{d}_2\} \quad (62)$$

Here we define two new variables given by:

$$\tilde{d}_k = \mu r_{kn} - r_{kt} \quad (63)$$

Using equations (21) and (28), we note:

$$\Delta \bar{v}_{Rel}^{CG} = \Delta v_{Reln}^{CG} \hat{n} + \Delta v_{Relt}^{CG} \hat{t}$$
$$= \Delta v_{Reln}^{CG} \cdot (\hat{n} + \mu \hat{t}), \quad (64)$$

and

$$|\Delta \bar{v}_{Rel}^{CG}| = |\Delta v_{Reln}^{CG}| \cdot \sqrt{1 + \mu^2}. \quad (65)$$

The ratio of tangent to normal relative velocity is again given by:

$$r = \frac{v_{Relt,i}^{PC}}{v_{Reln,i}^{PC}}.$$

Therefore we can write:

$$\bar{v}_{Rel,i}^{CG} = v_{Reln,i}^{CG} \cdot (\hat{n} + r\hat{t}). \quad (66)$$

Using equations (64) and (65), we can write the dot-product:

$$\bar{v}_{Rel,i}^{CG} \cdot \Delta \bar{v}_{Rel}^{CG} = v_{Reln,i}^{CG} \cdot \Delta v_{Reln}^{CG} \cdot (1 + \mu r) \quad (67)$$

Combining equations (55), (62), (65), and (67), we can rewrite (52) as follows:

$$(\Delta v_{Reln}^{CG})^2 \cdot \{A + \mu^2(1 + C) - 2\mu B\}$$
$$+ 2\Delta v_{Reln}^{CG} \cdot \{v_{Reln,i}^{CG}(1 + \mu r) + \omega_{1i}\tilde{d}_1 + \omega_{2i}\tilde{d}_2\}$$
$$+ \frac{2}{\bar{m}}E_{Loss} = 0, \quad (68)$$

Note equation (68) is a quadratic equation in the form:

$$a \cdot (\Delta v_{Reln}^{CG})^2 + b \cdot \Delta v_{Reln}^{CG} + c = 0,$$

where we identify:

$$a = A + \mu^2(1 + C) - 2\mu B$$

$$b = 2v_{Reln,i}^{CG}(1 + \mu r) + 2\omega_{1i}\tilde{d}_1 + 2\omega_{2i}\tilde{d}_2$$

$$c = \frac{2}{\bar{m}}E_{Loss}$$

Here we use the standard solution to the quadratic:

$$\Delta v_{Reln}^{CG} = \frac{-b \pm \sqrt{b^2 - 4ac}}{2a}$$

$$= \frac{-b}{2a} \times \left\{1 \pm \sqrt{1 - \frac{4ac}{b^2}}\right\}$$

or

$$\Delta v_{Reln}^{CG} = -\frac{(v_{Reln,i}^{CG}(1+\mu r) + \omega_{1i}\tilde{d}_1 + \omega_{2i}\tilde{d}_2)}{(A + \mu^2(1+C) - 2\mu B)} \times$$

$$\left\{1 \pm \sqrt{1 - \frac{4(A + \mu^2(1+C) - 2\mu B)\frac{2}{\bar{m}}E_{Loss}}{(v_{Reln,i}^{CG}(1+\mu r) + \omega_{1i}\tilde{d}_1 + \omega_{2i}\tilde{d}_2)^2}}\right\} \quad (69)$$

This is an exact expression for $\Delta v_{Reln}^{CG}$ as a function of the total energy lost to permanent crush and frictional effects. Note, this expression implicitly includes restitution effects as any restoration of crush energy would obviously reduce the value of $E_{Loss}$.

To be of practical use, a model of energy loss is needed which can naturally "factorize" or easily separate frictional-like effects from effects due to restoring forces. The approximation made in equation (35) allowed us to do just that. Rather than propose a model here that does this factorization, we will simply compare maximum energy loss estimates given by our two approaches to better understand what effect if any the equation (35) approximate has on our results.

First, we note here the expression under the radical must satisfy:

$$1 - \frac{4ac}{b^2} \geq 0$$

or

$$\frac{b^2}{4a} \geq c$$

Using this condition, we have the condition for maximum energy loss:

$$\frac{(v_{Reln,i}^{CG}(1+\mu r) + \omega_{1i}\tilde{d}_1 + \omega_{2i}\tilde{d}_2)^2}{(A + \mu^2(1+C) - 2\mu B)} \geq \frac{2}{\bar{m}}E_{Loss}$$

or

$$E_{Loss}^{Max} = \frac{1}{2}\bar{m}\frac{(v_{Reln,i}^{CG}(1+\mu r) + \omega_{1i}\tilde{d}_1 + \omega_{2i}\tilde{d}_2)^2}{(A + \mu^2(1+C) - 2\mu B)} \quad (70)$$

Now, let us assume there is no initial angular velocity for either vehicle. This simplifies equation (70) to:

$$E_{Loss}^{Max} = \frac{1}{2}\bar{m}(v_{Reln,i}^{CG})^2 \frac{(1+\mu r)^2}{(A + \mu^2(1+C) - 2\mu B)}$$

We know energy losses are maximized in the no-restitution limit. Looking at our expression for the approximate total energy loss, letting $\varepsilon = 0$, we have:

$$E_{Loss}^{Max} \approx E_{Loss}(\varepsilon = 0, \mu) = \frac{1}{2}\bar{m}(v_{Reln,i}^{PC})^2 \cdot q$$
$$\times \{1 + 2\mu r - \mu q[\mu(1+C) - B]\}$$

Suppose we now divide both of the above expressions by the kinetic energy term $\frac{1}{2}\bar{m}(v_{Reln,i}^{CG})^2$. We can then define the scale factors:

$$\gamma_{Exact} = \frac{(1+\mu r)^2}{(A + \mu^2(1+C) - 2\mu B)}$$

and

$$\gamma_{Approximate} = q \cdot \{1 + 2\mu r - \mu q[\mu(1+C) - B]\}.$$

We can relate these factors to the exact and approximate energy loss functions in the no restitution limit by:

$$E_{Loss}^{Max} = \gamma_{Exact} \times \frac{1}{2}\bar{m}(v_{Reln,i}^{CG})^2$$

and

$$E_{Loss}^{Max} \approx \gamma_{Approximate} \times \frac{1}{2}\bar{m}(v_{Reln,i}^{CG})^2$$

Figure 9 shows these two scale factors for our collision depicted in Figure 1. The scales factors are allowed to run over a full range of $\mu$. Note both the exact and approximate curves peak at 1.01. This is the value estimated for $\mu_{max}$ in our example calculations above. Figure 9 also shows a zoomed in plot with these two functions. We see relatively good agreement between the two, where here we have $r = 0.528$.

Figure 10 shows the scale factors versus $\mu$, where $r = 1, 10$, and $100$, where all other data is kept the same as in our example case. Again we observe exact matches in peak values at $\mu = \mu_{max}$. Also note, as $r$ increases, that is, as the impact becomes more side-swipe in nature, the peak shifts toward larger values of $\mu$. This has the obvious interpretation that as $r$ increases, the needed impulse ratio to fully retard relative tangential motion also increases. We also note as $r$ increases to large values, we see larger disagreement in the behavior of the curves themselves. This is particularly evident in the bottom panel of Figure 10. Generally however, such differences are not terribly important as a typical procedure is to simply estimate $\mu_{max}$ and scale it by some fractional value.

As a cross check, we allowed PC to fully vary for both vehicles along both axes. We also allowed restitution to vary between 0 and 1. We then calculated the values of $\gamma_{Exact}$ and $\gamma_{Approximate}$ at $\mu_{max}$ for 100,000 trials. Plotting the exact versus approximate scale factors gives us the perfect diagonal plot as shown in Figure 11, thereby serving as verification that the approximation works well in the no restitution limit.

**Conclusion**

We have presented a rigorous set of derivations arriving at the standard equations needed for planar impact dynamics equations commonly used in accident reconstruction applications. We have provided an example use case, where we have shown how to derive the sensitivity of the model outputs driven by variations to the model inputs. The energy factorization works well to model the dynamics of vehicle collisions, though there can be large variations in the exact behavior of energy loss versus impulse ratio. Such differences prove unimportant if the model is used correctly.

*Bob Scurlock, Ph. D. is a Research Associate at the University of Florida, Department of Physics and works as a consultant for the accident reconstruction and legal community. He can be reached at BobScurlockPhD@gmail.com. His website offering free analysis software can be found at: ScurlockPhD.com.*

*James Ipser, Ph. D. is Professor Emeritus at the University of Florida, Department of Physics. He regularly consults and provides expert opinion in the areas of vehicular accident reconstruction and biomechanical physics. He can be reached at JIpser@gmail.com. His website can be found at JIpsier.com.*

**Appendix**

Here we consider an alternative definition of restitution, given by ratio of final-to-initial closing-velocities at the CG:

$$(\varepsilon^{CG})^2 = \frac{|\bar{v}_{Rel,f}^{CG}|^2}{|\bar{v}_{Rel,i}^{CG}|^2} \quad (A1)$$

From simple vector addition, we know:

$$\bar{v}_{Rel,f}^{CG} = \bar{v}_{Rel,i}^{CG} + \Delta\bar{v}_{Rel}^{CG}$$

This implies:

$$|\bar{v}_{Rel,f}^{CG}|^2 = (\varepsilon^{CG})^2 \cdot |\bar{v}_{Rel,i}^{CG}|^2$$
$$= |\bar{v}_{Rel,i}^{CG}|^2 + |\Delta\bar{v}_{Rel}^{CG}|^2 + 2\bar{v}_{Rel,i}^{CG} \cdot \Delta\bar{v}_{Rel}^{CG}$$

or

$$|\Delta\bar{v}_{Rel}^{CG}|^2 + 2\bar{v}_{Rel,i}^{CG} \cdot \Delta\bar{v}_{Rel}^{CG} + [1 - (\varepsilon^{CG})^2] \cdot |\bar{v}_{Rel,i}^{CG}|^2 = 0 \quad (A2)$$

From equation (64) we have:
$$\Delta\bar{v}_{Rel}^{CG} = \Delta v_{Reln}^{CG}\hat{n} + \Delta v_{Relt}^{CG}\hat{t}$$
$$= \Delta v_{Reln}^{CG} \cdot (\hat{n} + \mu\hat{t})$$

and from equation (67):

$$\bar{v}_{Rel,i}^{CG} \cdot \Delta\bar{v}_{Rel}^{CG} = v_{Reln,i}^{CG} \cdot \Delta v_{Reln}^{CG} \cdot (1 + \mu r)$$

Using equations (64) and (67), equation (A2) can be re-expressed as:

$$(1+\mu^2)(\Delta v_{Reln}^{CG})^2 + 2v_{Reln,i}^{CG} \cdot \Delta v_{Reln}^{CG} \cdot (1+\mu r)$$
$$+[1-(\varepsilon^{CG})^2] \cdot (1+r^2)|\bar{v}_{Reln,i}^{CG}|^2 = 0$$

From equation (68), we have an exact form for the quadratic which is given by:

$$(\Delta v_{Reln}^{CG})^2 \cdot \{A + \mu^2(1+C) - 2\mu B\}$$
$$+2\Delta v_{Reln}^{CG} \cdot \{v_{Reln,i}^{CG}(1+\mu r) + \omega_{1i}\tilde{d}_1 + \omega_{2i}\tilde{d}_2\}$$
$$+\frac{2}{\bar{m}}E_{Loss} = 0$$

In order for the equations (A2) and (68) to be consistent, they must share the same solutions for $\Delta v_{Reln}^{CG}$. To simplify our analysis here, let us assume that the contact force is only along the normal axis. This implies $r = \mu = 0$. Let's also assume $\omega_{1i} = \omega_{2i} = 0$. Now the two equations which must yield the same quadratic solutions are:

$$(\Delta v_{Reln}^{CG})^2 + 2v_{Reln,i}^{CG} \cdot \Delta v_{Reln}^{CG} + [1-(\varepsilon^{CG})^2] \cdot |\bar{v}_{Reln,i}^{CG}|^2 = 0$$

and

$$(\Delta v_{Reln}^{CG})^2 \cdot \{A\} + \Delta v_{Reln}^{CG} \cdot \{2v_{Reln,i}^{CG}\} + \frac{2}{\bar{m}}E_{Loss} = 0$$

Subtracting the two equations, we have

$$(\Delta v_{Reln}^{CG})^2 \cdot \{A-1\}$$
$$+\left\{\frac{2}{\bar{m}}E_{Loss} - [1-(\varepsilon^{CG})^2] \cdot |\bar{v}_{Reln,i}^{CG}|^2\right\} = 0$$

Using $(\Delta v_{Reln}^{CG})^2 = (1+\varepsilon^{CG})^2 \cdot |\bar{v}_{Rel,i}^{CG}|^2$ this becomes:

$$(1+\varepsilon^{CG})^2 \cdot |\bar{v}_{Rel,i}^{CG}|^2 \cdot \{A-1\}$$
$$+\left\{\frac{2}{\bar{m}}E_{Loss} - [1-(\varepsilon^{CG})^2] \cdot |\bar{v}_{Reln,i}^{CG}|^2\right\} = 0$$

or

$$(1+2\varepsilon^{CG}+(\varepsilon^{CG})^2) \cdot \{A-1\}$$
$$+\frac{2}{\bar{m}}E_{Loss}/|\bar{v}_{Rel,i}^{CG}|^2 - 1 + (\varepsilon^{CG})^2 = 0$$

Now $\varepsilon^{CG}$ must be solved for using the quadratic:

$$(\varepsilon^{CG})^2\{A\} + 2\varepsilon^{CG}\{A-1\}$$
$$+\left\{A-2+\frac{2}{\bar{m}}E_{Loss}/|\bar{v}_{Rel,i}^{CG}|^2\right\} = 0$$

Let us take as a special case the full restitution limit $\frac{2}{\bar{m}}E_{Loss} = 0$. Then our equation simplifies to:

$$(1+\varepsilon^{CG})^2 \cdot |\bar{v}_{Rel,i}^{CG}|^2 \cdot \{A-1\}$$
$$+\left\{-[1-(\varepsilon^{CG})^2] \cdot |\bar{v}_{Reln,i}^{CG}|^2\right\} = 0$$

or

$$(1+\varepsilon^{CG})^2 \cdot \{A-1\} - (1+\varepsilon^{CG})(1-\varepsilon^{CG}) = 0$$

This can be further simplified to:

$$(1+\varepsilon^{CG}) \cdot \{A-1\} - (1-\varepsilon^{CG}) = 0$$

Solving for $(\varepsilon^{CG})$ we finally have:

$$(\varepsilon^{CG}) = (2-A)/A$$

or

$$(\varepsilon^{CG}) = \frac{1 - \bar{m}\left(\frac{r_{1t}^2}{I_1} + \frac{r_{2t}^2}{I_2}\right)}{1 + \bar{m}\left(\frac{r_{1t}^2}{I_1} + \frac{r_{2t}^2}{I_2}\right)}$$

Returning to equation (A2), we may wish to write a general expression for $|\Delta \bar{v}_{Rel}^{CG}|$ by solving the quadratic form. Here we make the associations:

$$a = 1$$
$$b = 2\bar{v}_{Rel,i}^{CG} \cdot \Delta \hat{v}_{Rel}^{CG} = 2|\bar{v}_{Rel,i}^{CG}|\cos\varphi$$
$$c = [1-(\varepsilon^{CG})^2] \cdot |\bar{v}_{Rel,i}^{CG}|^2$$

where $\varphi$ defines the angle between the vectors $\Delta \bar{v}_{Rel}^{CG}$ and $\bar{v}_{Reln,i}^{CG}$. The solution is given by:

$$|\Delta \bar{v}_{Rel}^{CG}| = \frac{-b \pm \sqrt{b^2 - 4ac}}{2a}$$

This implies:

$$|\Delta \bar{v}_{Rel}^{CG}|$$
$$= \frac{-2|\bar{v}_{Rel,i}^{CG}|\cos\varphi \pm \sqrt{\left(2|\bar{v}_{Rel,i}^{CG}|\cos\varphi\right)^2 - 4[1-(\varepsilon^{CG})^2] \cdot |\bar{v}_{Rel,i}^{CG}|^2}}{2}$$

Finally, the above simplifies to:

$$|\Delta \bar{v}_{Rel}^{CG}| = -|\bar{v}_{Rel,i}^{CG}|\left\{\cos\varphi \pm \sqrt{(\varepsilon^{CG})^2 - (\sin\varphi)^2}\right\}$$

Here we make two observations. First, note $\varepsilon^{CG}$ does not have to be positive, as the above solution is invariant under the transformation $\varepsilon^{CG} \to -\varepsilon^{CG}$. Second, we kept the general solution including both roots. We can easily explore the implication of this as follows. Suppose we let $\varphi = \pi$, that is, the case in which the change-in-relative-velocity is exactly anti-parallel to the initial relative velocity. This of course will yield:

$$|\Delta \bar{v}_{Rel}^{CG}| = -|\bar{v}_{Rel,i}^{CG}|\{-1 \pm |\varepsilon^{CG}|\}$$
$$= |\bar{v}_{Rel,i}^{CG}| \mp |\varepsilon^{CG}| \cdot |\bar{v}_{Rel,i}^{CG}|$$

Here we see the "+" root will tend to cause $|\Delta \bar{v}_{Rel}^{CG}| \leq |\bar{v}_{Rel,i}^{CG}|$ where as the "$-$" root will tend to cause $|\Delta \bar{v}_{Rel}^{CG}| \geq |\bar{v}_{Rel,i}^{CG}|$. The former solution is consistent with a two objects whose relative velocity projected on the force axis remains pointing along the same direction through reduced in magnitude. The latter solution is consistent with the projected relative velocity changing directions.

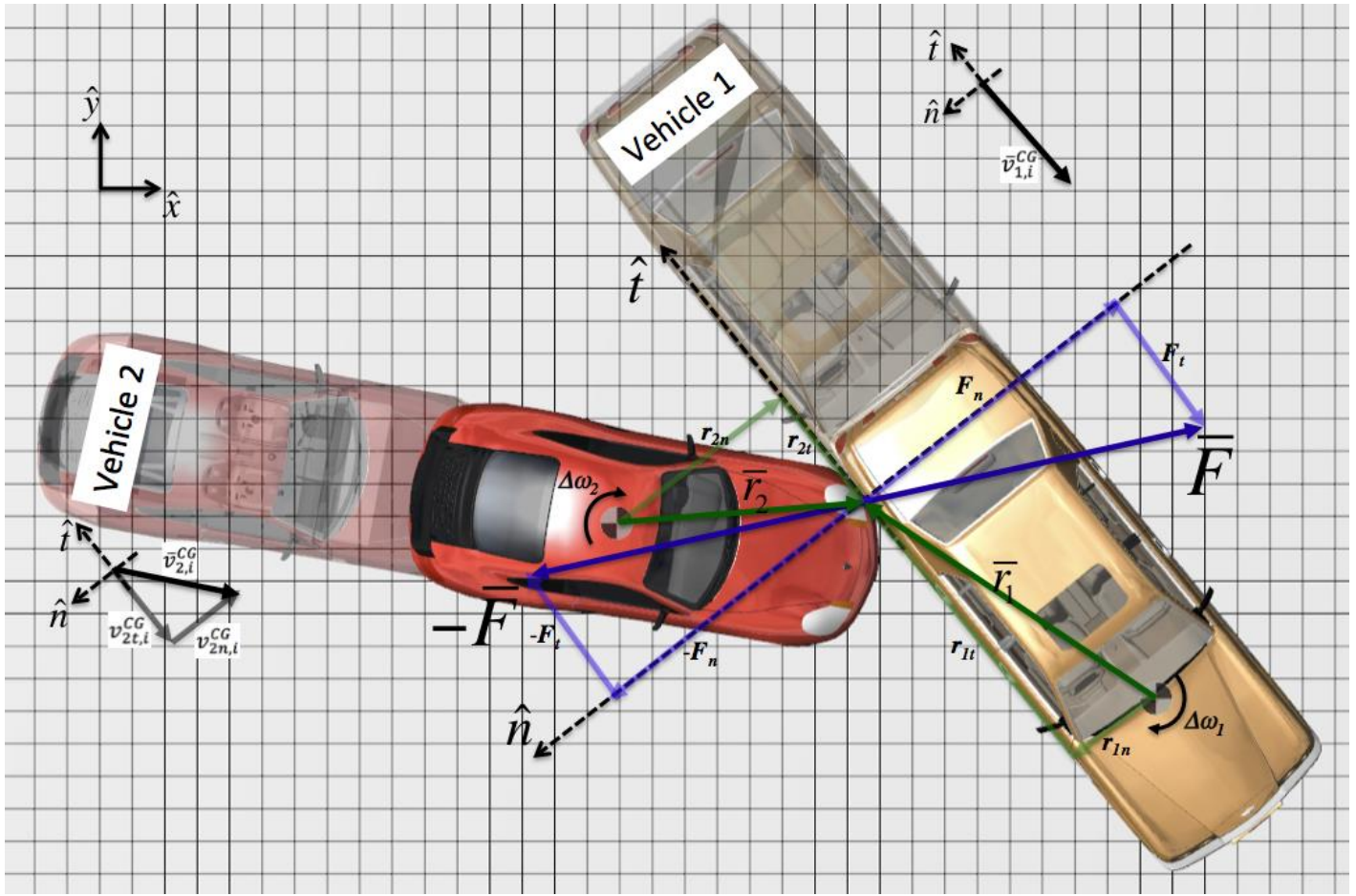

**Figure 1:** Aerial view of side impact collision. Definitions of normal and tangent axes are shown, as well as the decompositions of lever-arm and force vectors.

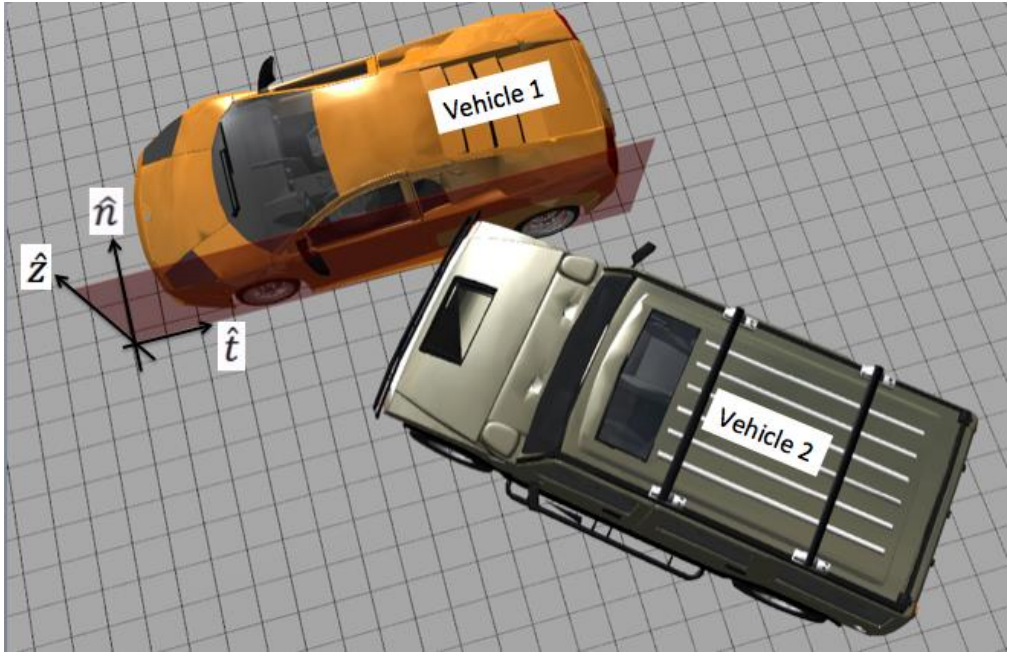

**Figure 2:** Diagram showing example impact orientation of two vehicles. The normal axis is pointing anti-parallel to the surface vector normal to the driver side of Vehicle 1. The tangent axis is running from front to rear of Vehicle 1.

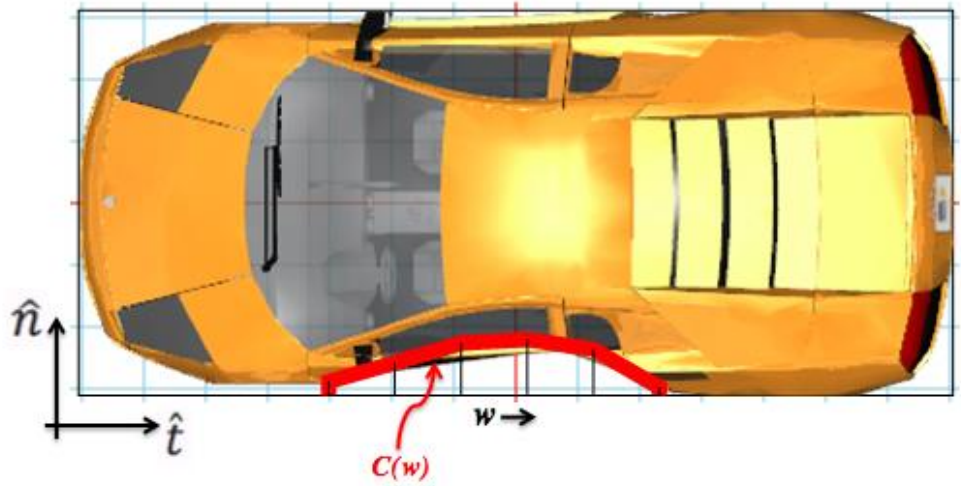

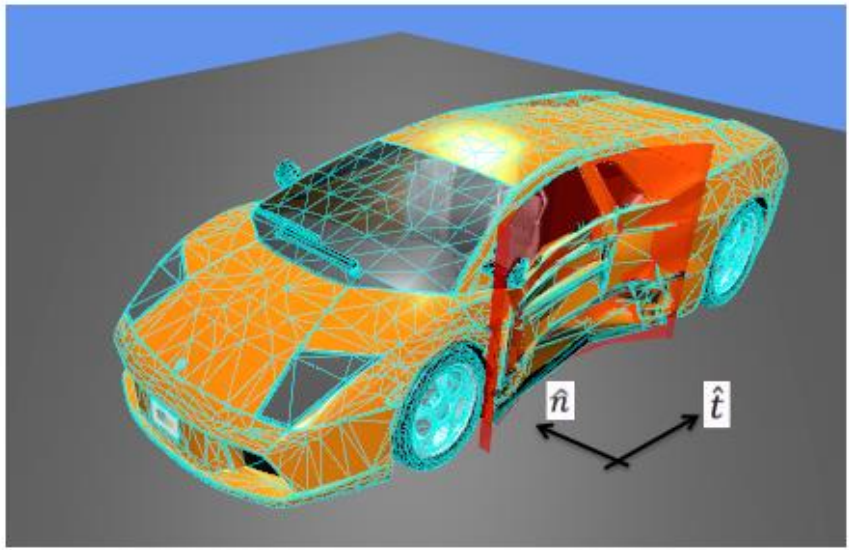

**Figure 3: Example crush profile shown for impact depicted in Figure 2.**

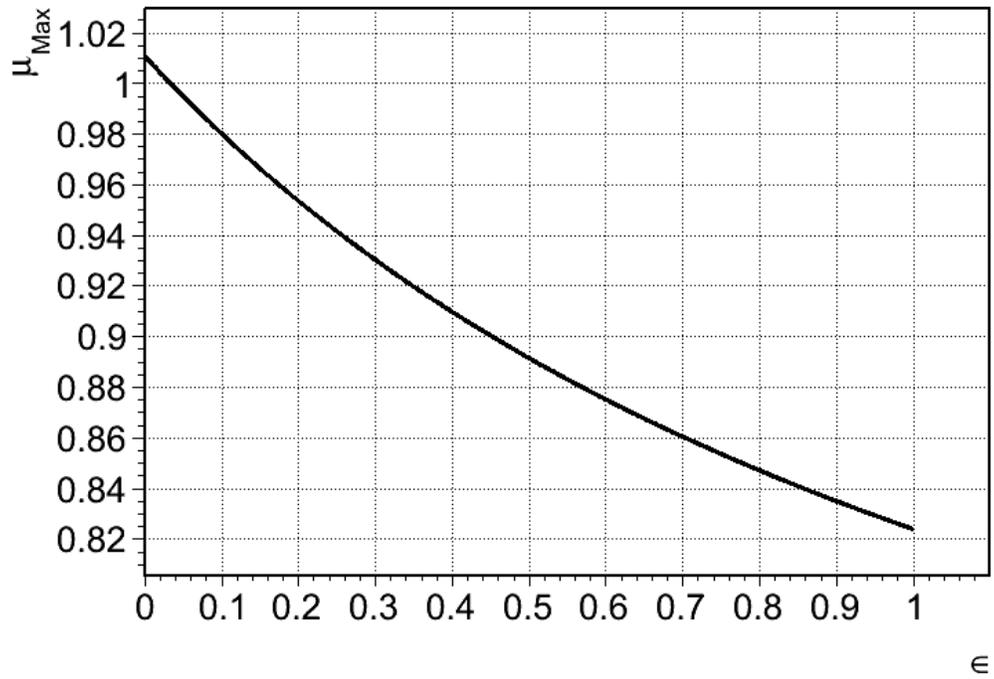

Figure 4: Plot showing relationship between coefficient-of-restitution and maximum impulse ratio for the collision shown in Figure 1.

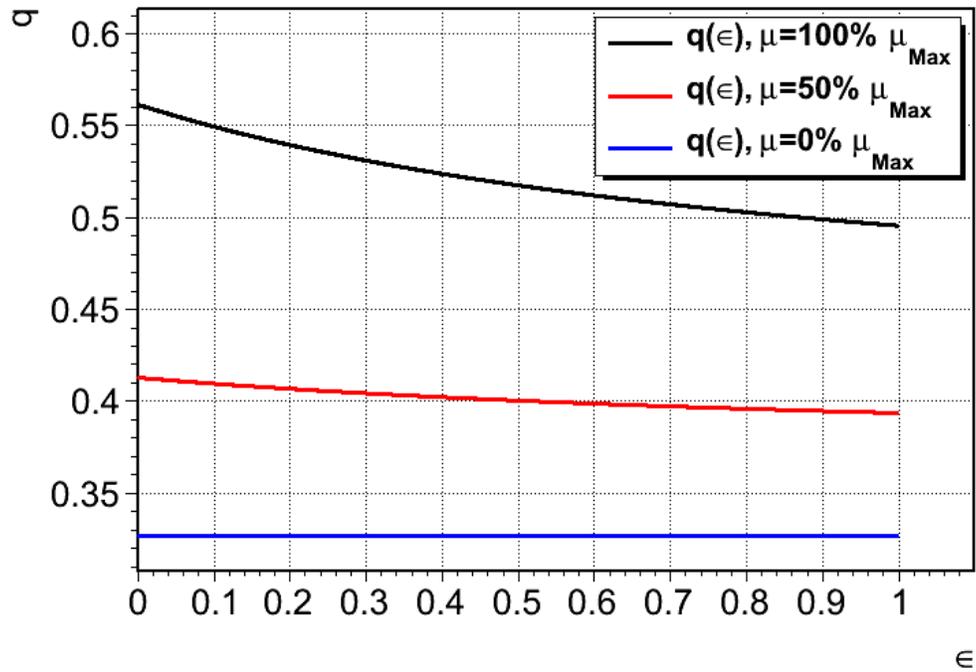

**Figure 5: Variation of *q* shown as a function of coefficient of restitution. Three scenarios for the impulse ratio are shown.**

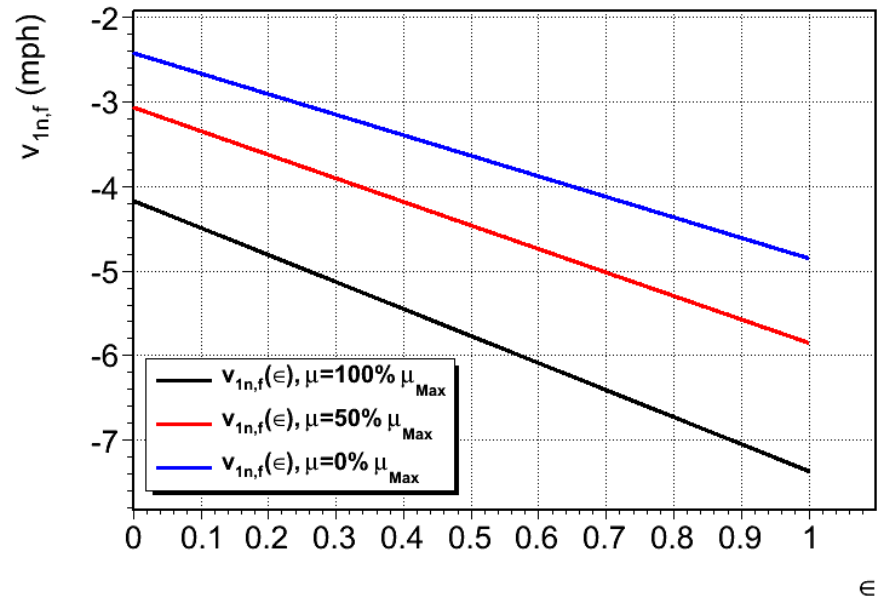

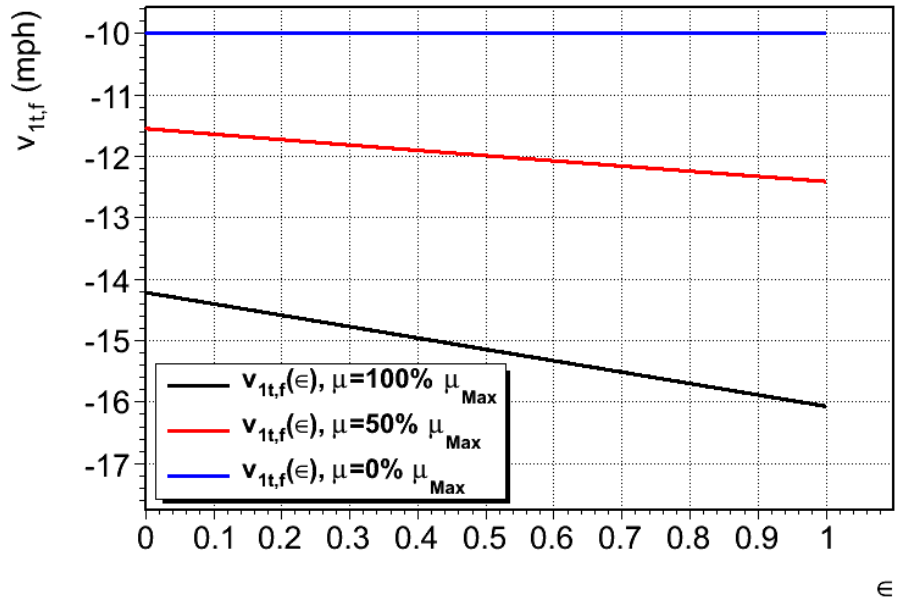

**Figure 6: Normal (top) and tangent (bottom) final velocity components for vehicle 1 shown versus the impulse ratio and restitution.**

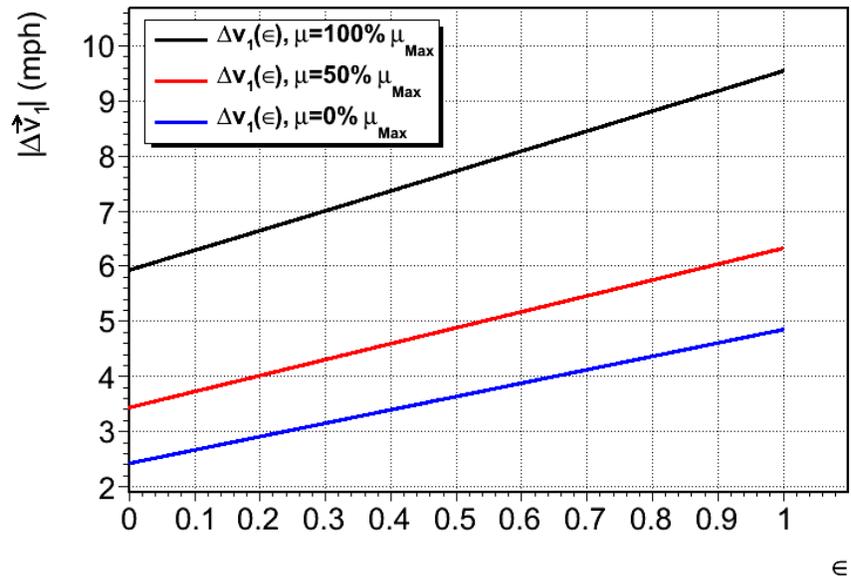

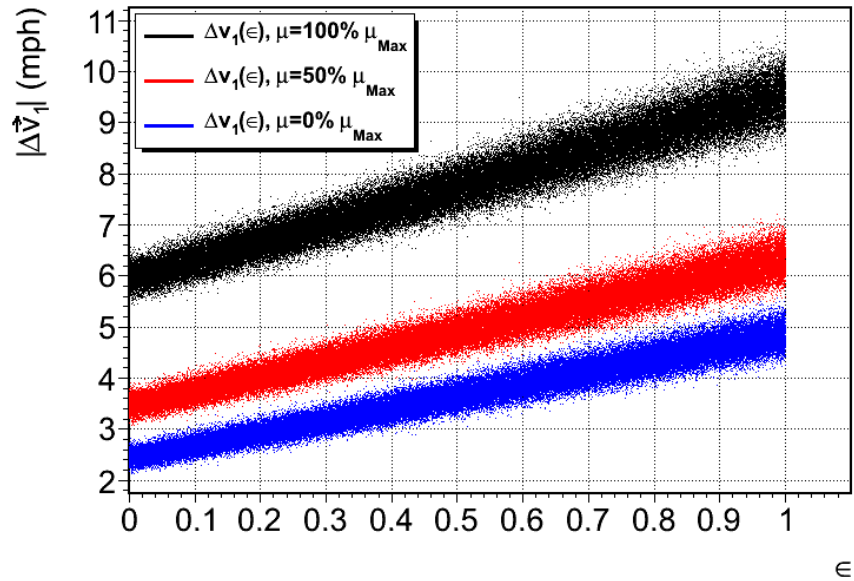

**Figure 7**: Top: Delta-V(Vehicle1) versus impulse ratio and restitution for collision depicted in Figure 1. Bottom: Delta-V(Vehicle1) allowing for variation in *PC* along the tangent direction.

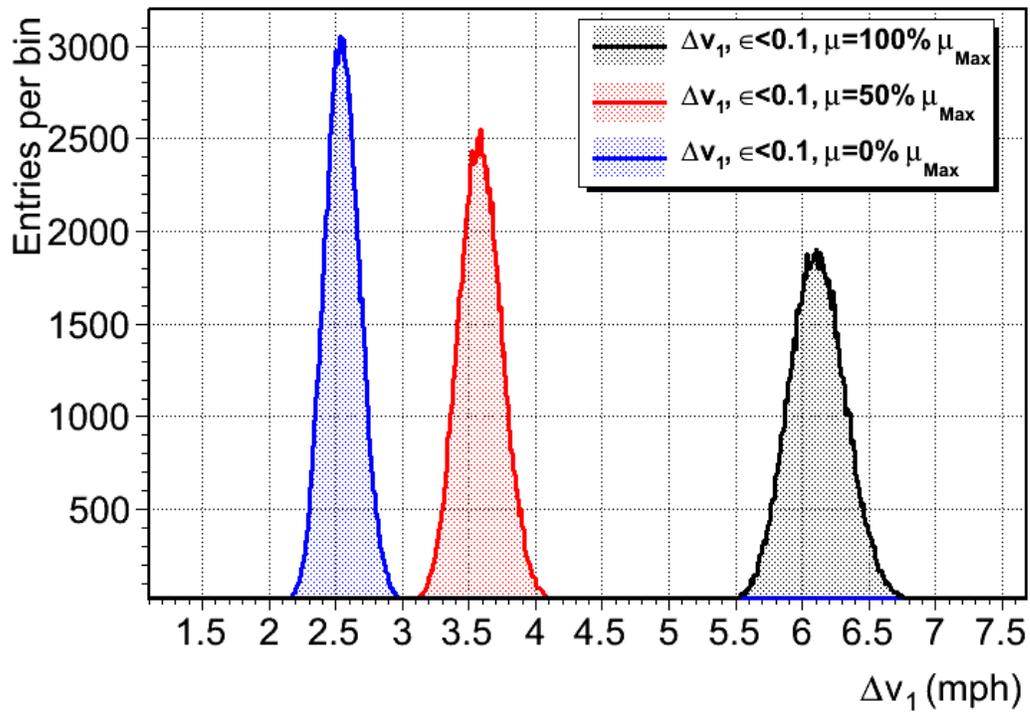

**Figure 8: Variation in Delta-V(Vehicle1) for restitution less than 0.1, where $r_{1t}$ is smeared by 0.5 ft about its central value.**

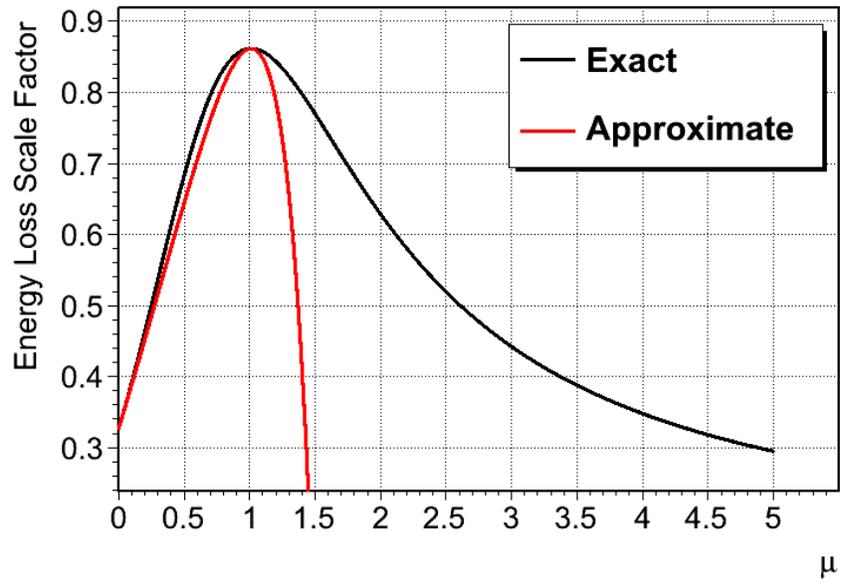

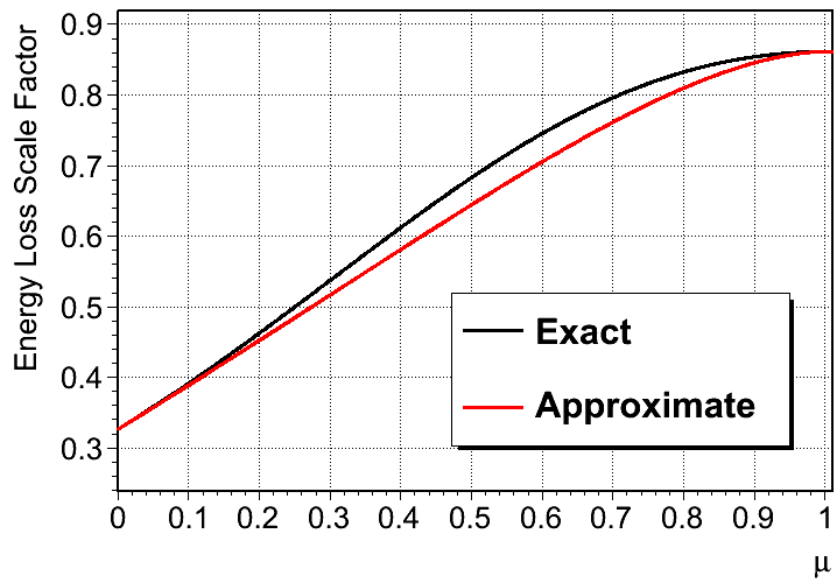

**Figure 9: Top:** Energy loss scale factors versus impulse ratio for our impact depicted in Figure 1, where *r*=0.528. The bottom plot shows the same as above, focused in on the region $\mu = 0$ to 1.

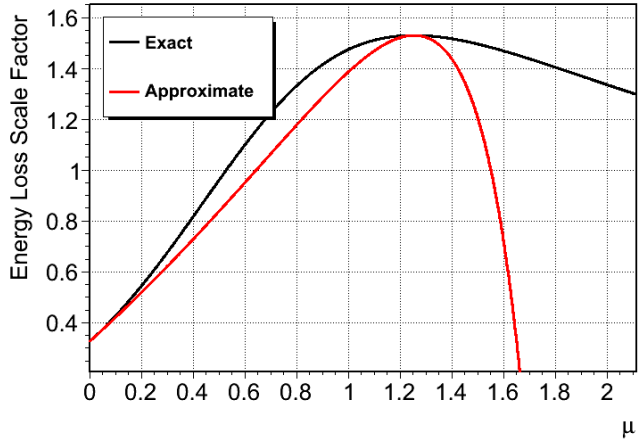
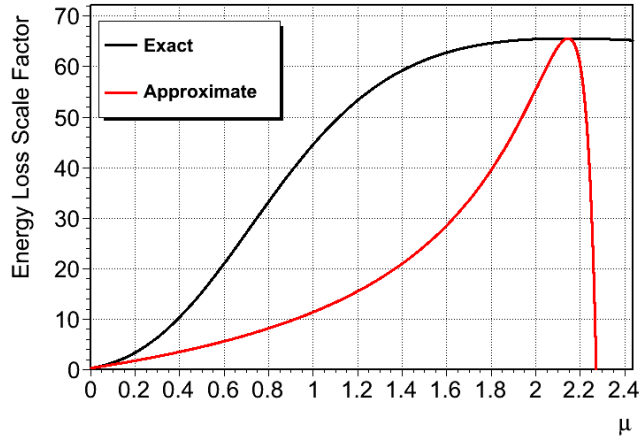
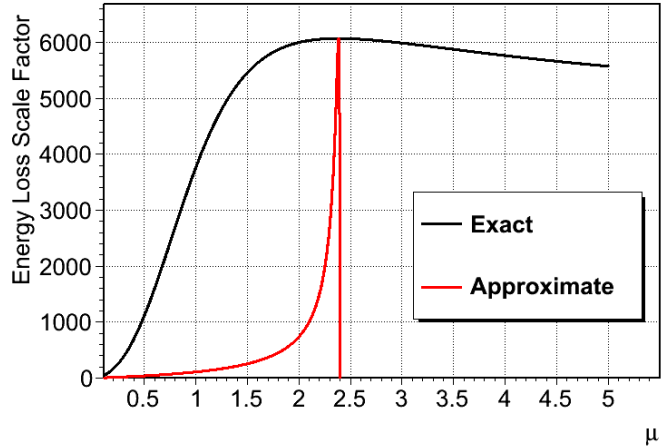

**Figure 10: Energy loss scale factors using example collision data, but with r=1 (top), r=10 (middle), and r=100 (bottom).**

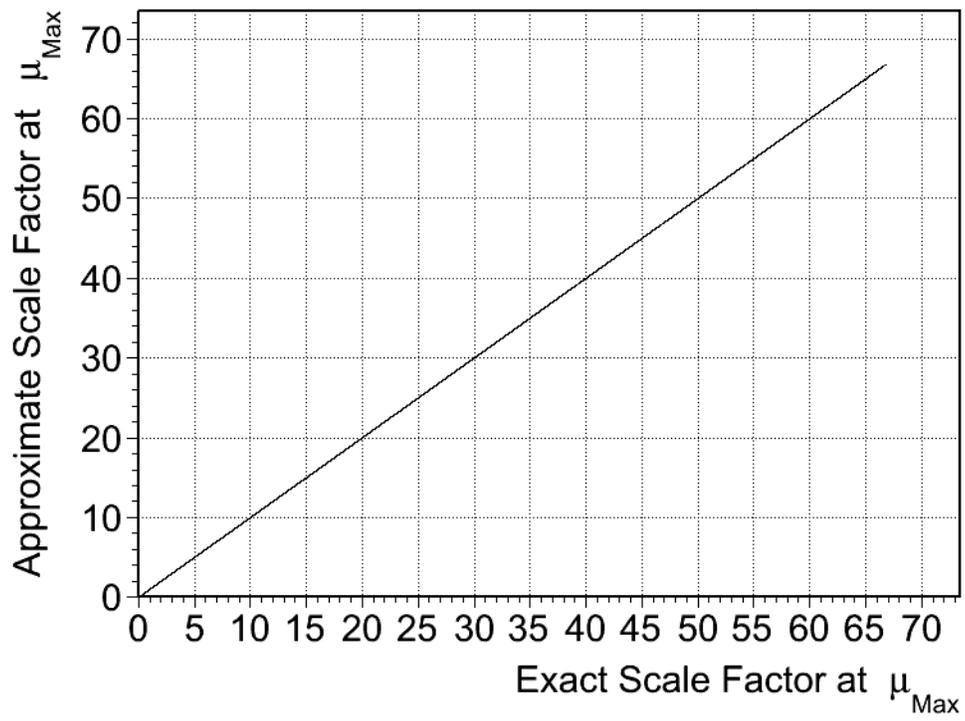

Figure 11: Approximate versus exact energy loss scale factor for randomly varying input conditions. Perfect agreement is observed.